\documentclass[sigconf]{acmart}
\renewcommand\footnotetextcopyrightpermission[1]{} % removes footnote with conference information in first column
\AtBeginDocument{%
  }

\usepackage{marginnote}
\newcommand{\pageenlarge}[1]{\enlargethispage{#1\baselineskip}}

\usepackage{tcolorbox}
\tcbuselibrary{breakable}
\usepackage{multirow}
\usepackage{makecell}

\begin{document}

\title{GLoSS: Generative Language Models with Semantic Search for Sequential Recommendation}

\author{Krishna Acharya}
\affiliation{%
  \institution{Georgia Institute of Technology}
  \city{Atlanta}
  \state{Georgia}
  \country{USA}
}
\email{kacharya33@gatech.edu}

\author{Aleksandr V. Petrov}
\affiliation{
  \institution{Viator, Tripadvisor}
  \city{Glasgow}
  \country{UK}}
\email{spetrov@tripadvisor.com}

\author{Juba Ziani}
\affiliation{
  \institution{Georgia Institute of Technology}
  \city{Atlanta}
  \state{Georgia}
  \country{USA}}
\email{jziani3@gatech.edu}

\renewcommand{\shortauthors}{Acharya et al.}

\begin{abstract} 
We propose \textbf{\underline{G}enerative \underline{Lo}w-rank language model with \underline{S}emantic \underline{S}earch (GLoSS)}, a generative recommendation framework that combines large language models with dense retrieval for sequential recommendation. Unlike prior methods such as GPT4Rec, which rely on lexical matching via BM25, GLoSS uses semantic search to retrieve relevant items beyond lexical matching. For query generation, we employ 4-bit quantized LlaMA-3 models fine-tuned with low-rank adaptation (LoRA), enabling efficient training and inference on modest hardware. We evaluate GLoSS on three real-world Amazon review datasets: Beauty, Toys, and Sports, and find that it achieves state-of-the-art performance. Compared to traditional ID-based baselines, GLoSS improves Recall@5 by 33.3\%, 52.8\%, and 15.2\%, and NDCG@5 by 30.0\%, 42.6\%, and 16.1\%, respectively. It also outperforms LLM-based recommenders such as P5, GPT4Rec, LlamaRec and E4SRec with Recall@5 gains of 4.3\%, 22.8\%, and 29.5\%. Additionally, user segment evaluations show that GLoSS performs particularly well for cold-start users in the Amazon Toys and Sports datasets, and benefits from longer user histories in Amazon Beauty dataset, demonstrating robustness across different levels of interaction lengths.
Our code and model checkpoints are publicly available at:
\url{https://github.com/krishnacharya/GLoSS}
\end{abstract}

\maketitle

\begin{figure}
\includegraphics[width=0.5\textwidth]{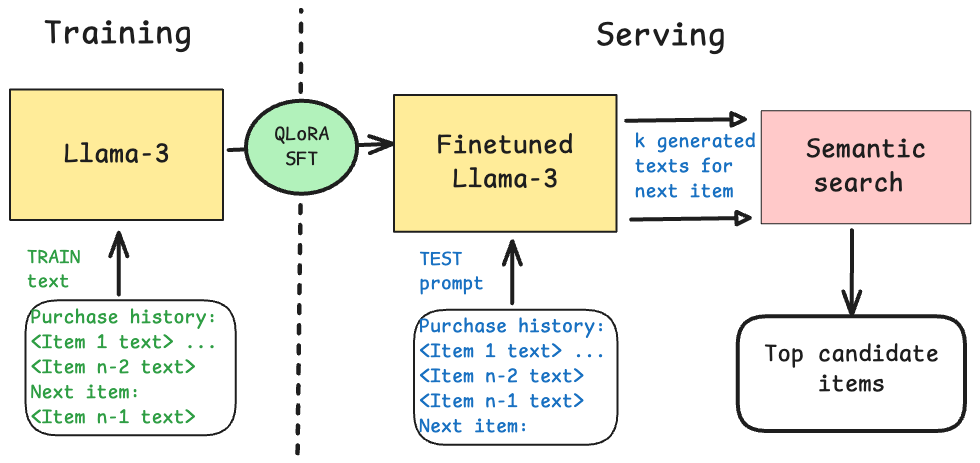}
\caption{GLoSS: a finetuned LLM generates queries, which are then used to perform semantic search over the item catalog.}
\label{fig:gloss}
\end{figure}

\pageenlarge{2}
\section{Introduction}
\looseness -1 Sequential Recommender Systems predict the next item in the sequence of user-to-item interaction. Until recently, \emph{id-based} approaches dominated the field of sequential recommendation: these methods typically learn an embedding representation for every item in the catalog, as well as a \emph{sequence encoder}---a model that represents the whole sequence of user-item interactions in the same space of embeddings. Id-based methods then compute item scores using a similarity function (e.g., dot product) between the sequence embedding and the item embedding. Some of the most notable examples of id-based methods include SASRec~\cite{sasrec} and BERT4Rec~\cite{bert4rec}.

\looseness -1 Despite achieving high recommendation accuracy on many sequential datasets, id-based methods have a number of limitations. In particular, these id-based methods require item embeddings to be \emph{learned} for every item in the catalog, meaning that new items can be only recommended after the recommendation model has been retrained. In addition, models like SASRec and BERT4Rec are trained only using data directly presented in the training set, which constrains their ability to generalize beyond seen patterns or infer relevance based on rich item metadata or natural language descriptions. 

In contrast, textual document retrieval systems do not suffer from these limitations: documents (equivalent of ``items'' in recommender systems) are typically indexed independently of the scoring model, and, hence can be retrieved immediately after being added to index. 
Moreover, the widespread adaptation of pre-trained language models (such as BERT~\cite{BERT}), has enabled fine-tuning for retrieval, allowing the models to generalize beyond the training data and retrieve relevant documents even in the absence of direct lexical overlap, leading to  substantial improvements in retrieval effectiveness~\cite{karpukhinDensePassageRetrieval2020,reimersSentenceBERTSentenceEmbeddings2019}. 

\looseness -1 Recent advances in generative language models have enabled their integration into recommendation pipelines. For instance, P5Rec~\cite{P5Rec} adopts T5~\cite{T5paper} as a unified backbone to support multiple tasks including sequential recommendation, rating prediction, and review summarization. GPT4Rec~\cite{GPT4Rec} uses GPT-2 to generate multiple queries per user, which are then matched to items via a BM25 lexical retriever. More recently, LlamaRec~\cite{Llamarec} employs LLaMA-2 as a reranker over candidates selected by an LRU-based recommender~\cite{LRURec}. Other methods, such as CALRec and GenRec~\cite{Calrec,Genrec}, follow a similar query generation and retrieval framework, combining LLMs like PaLM~\cite{anil2023palm} or LLaMA-2~\cite{llama2} with BM25-based lexical search.

However, many of these methods use BM25, a lexical matching algorithm that does not capture the semantic meaning of generated queries. Furthermore, approaches like P5, GPT4Rec, and CALRec rely on older language models (e.g., T5, GPT-2, PaLM) and use full fine-tuning, which is computationally intensive.

Over the past five years, generative language models have improved significantly due to advancements in architecture, better training data, and more effective training methods. This progress has resulted in models like the LLaMA-3 series~\cite{llama3}, which are substantially more capable for tasks such as query generation. Additionally, parameter-efficient fine-tuning techniques such as QLoRA~\cite{qlora} have dramatically reduced the resource requirements for training, enabling efficient fine-tuning on consumer GPUs.

\pageenlarge{2}
Inspired by the GPT4Rec approach and aiming to address its limitations, we propose \textbf{\underline{G}enerative \underline{Lo}w-rank language model with \underline{S}emantic \underline{S}earch (GLoSS)}, a generative recommendation framework for sequential recommendation. Unlike GPT4Rec, GLoSS uses semantic search (aka. dense retrieval) to enable item recommendation beyond direct lexical matching. Our query generation backbone uses 4-bit quantized LLaMA-3 models fine-tuned via low-rank adapters.

\looseness -1 We evaluate GLoSS on three real-world Amazon review datasets: \\Beauty, Toys, and Sports \cite{amzn2014}---and demonstrate that it achieves state-of-the-art performance. Compared to ID-based baselines, GLoSS yields substantial gains in Recall@5 (33.3\% on Beauty, 52.8\% on Toys, and 15.2\% on Sports) and NDCG@5 (30.0\%, 42.6\%, and 16.1\%, respectively). GLoSS also outperforms LLM-based recommenders such as P5, GPT4Rec, LlamaRec and E4SRec, with Recall@5 improvements of 4.3\% on Beauty, 22.8\% on Toys, and 29.5\% on Sports—while maintaining competitive NDCG@5 without relying on a reranking stage.

In summary, the contributions of this paper are as follows
\begin{enumerate}
    \item We apply low-rank adaptation (LoRA) to fine-tune 4-bit quantized LLaMA-3 models for query generation, showing that high-quality generative performance can be achieved with minimal resources.
    \item We integrate dense retrieval into the generative recommendation pipeline to enable retrieval beyond lexical matching. Our analysis in Section~\ref{sec:RQ-DenseRetSize} shows that dense retrieval is a critical component, yielding substantial gains in NDCG@5 and Recall@5 over BM25.
    \item Through extensive experiments on Amazon datasets, we show that GLoSS outperforms both ID- and LLM-based baselines with consistent double-digit improvements in Recall@5 and NDCG@5.
    \item We analyze GLoSS across user segments based on interaction history length: cold-start, regular, and power users. To our knowledge, this is the first such study for LLM-based recommenders. GLoSS performs best on cold-start users in Toys and Sports, and benefits from longer histories in Beauty, demonstrating robustness across user interaction lengths.
\end{enumerate}
The rest of the paper is organised as follows: Section~\ref{sec:related} contains the description of similar methods in RecSys and IR; Section~\ref{sec:methodology} describes our proposed GLoSS; Section~\ref{sec:experiments} describes experimental evaluation; Section~\ref{sec:concl} contains concluding remarks. We now describe prior related to ours.

\section{Recommendation through Retrieval} \label{sec:related}
This section describes methods from the fields of Recommender Systems and Information Retrieval. Section~\ref{ssec:content} describes classic content-based methods; Section~\ref{ssec:genrec} describes recent generative methods; Section~\ref{ssec:querygen} discusses similar methods from the IR field.

\pageenlarge{3}
\subsection{Content-based filtering}\label{ssec:content}
\label{Sec:relatedworks-Content-based}
    Our work retrieves items from the catalog using textual description, generated by a language model. Hence, our method has similarities with \emph{content-based} recommendation methods that were developed in the early days of the recommender systems research. Indeed, in the mid-1990s, Lang~\cite{LANG1995331} developed a news recommendation algorithm that used textual representation of past user netnews articles to recommend future articles to read, using lexical matching. Pazzani et al., developed the ``Syskill \& webert'' software agent~\cite{PAZANI1996} that generated a search query to find interesting websites based on past visited websites. Balabanovi{\'c} et al.\ developed a famous ``FAB'' recommender system that combined textual matching with id-based collaborative filtering. While these early content-based methods resemble certain similarities with our proposed method, they have relied on simplistic natural language methods,  such as TF-IDF~\cite{sparck1972statistical} vector space model. Due to the simplicity of these models, since the early 2000s, the id-based methods (e.g. Matrix Factorization~\cite{koren2009matrix}) showed better performance on many recommendation tasks and since then they remain de-facto standard. However, given the rapid progress of Large Language Models within the the last few years, we argue that a new generation of content-based models can make an unexpected comeback. Indeed, our proposed GLoSS can be seen as a content-based method, as it does not learn item representations and only uses their textual representations for matching. Nevertheless, in our experiments it achieves superior performance compared to state-of-the-art ID-based methods, such as SASRec, TIGER and ActionPiece.

\subsection{Generative Sequential Recommendation} \label{ssec:genrec}
Recently, a number of works used generative models to directly generate item id. Some of the most notable examples include HSTU~\cite{zhai2024actions} that used atomic item ids,  P5~\cite{P5Rec} that learns to generate numeric item IDs represented as strings, TIGER~\cite{TIGER} that uses \emph{semantic IDs} obtained from item embeddings, GPTRec\footnote{not to be confused with GPT\textbf{4}Rec}~\cite{GPTRec,petrovAligningGPTRecBeyondAccuracy2024} that obtains item ids from collaborative item embeddings, ActionPiece~\cite{ActionPiece} that constructs item ids from unordered set of its features. A central challenge addressed by these works is designing item id representations that generative models can effectively learn and reproduce. However, regardless of how item ids are represented, large language models have not encountered these ids during pre-training (as they are specific for the method). As a result, the models must learn to generate item ids from scratch, without leveraging the vast internet-scale training data that typically benefits other text generation tasks.

\looseness -1 In contrast, in our GLoSS—similarly to GPT4Rec~\cite{GPT4Rec}, which served as our inspiration—a large language model is used to generate textual queries for a retrieval system. That is, we employ the LLM in its native textual domain, without expanding its vocabulary or requiring it to learn atypical token sequences. This design allows GLoSS to fully leverage the internet-scale corpus used during the model’s pretraining.

\pageenlarge{2}
\subsection{Automatic Query Generation in Information Retrieval}\label{ssec:querygen}
Our approach uses an LLM to formulate queries which are then sent to a retrieval system in order to retrieve relevant items. This is similar to automatic query generation approaches in Information Retrieval. For example, Cui et al.~\cite{cui2003query} used search terms mined from user's log data to expand search queries. Jones et al.~\cite{jones2006Substitions} developed a statistical model that returned new queries that could be used as the substitutions to original queries. Recently, LLMs have also been employed  for query generation. For example, Ye et al.~\cite{ye2023enhancing}. used LLMs to rewrite the original query in order to retrieve better search results. While these methods have some similarities with our method, they are focused on the classic information retrieval task, where user provides an explicit query. In contrast, our approach focuses on recommendation task, that does not require user to provide any explicit inputs. 

With this, we conclude the overview of related work. In summary, no prior work uses a large language model to generate natural language queries for semantic item retrieval in a recommendation setting. The most similar approach to ours is GPT4Rec~\cite{GPT4Rec}, which relies on lexical matching rather than dense retrieval; we provide a detailed comparison between GLoSS and GPT4Rec in Section~\ref{sec:GPT,LlamrecvsGloss}. We now describe details of GLoSS methodology.

\section{GLoSS Methodology} \label{sec:methodology}
In this section, we describe our methodology.  Section~\ref{ssec:seqrec} formalizes the sequential recommendation task. Section~\ref{sec:lis-baseline} describes LIS, a simple training-free content-based model we use as a baseline. Section~\ref{sec:Gloss-pipeline} describes our GLoSS pipeline. Section~\ref{sec:GPT,LlamrecvsGloss} compares GloSS with the most related methods, GPT4Rec and LlamaRec.

\subsection{Sequential recommendation} \label{ssec:seqrec}
Assume that we have a set of users and items, denoted by $\mathcal{U}$ and $\mathcal{I}$, respectively, where $u \in \mathcal{U}$ denotes a user and $i \in \mathcal{I}$ denotes an item. The numbers of users and items are denoted as $|\mathcal{U}|$ and $|\mathcal{I}|$, respectively. A user $u$ is represented by a chronologically-ordered interaction sequence with items $\{i_1, \dots, i_{n-1}\}$, where $n-1$ is the number of interactions and $i_t$ is the $t$-th item that the user $u$ has interacted with. For convenience, we use $i_{j:k}$ to denote the subsequence, i.e., $i_{j:k} = \{i_j, \dots, i_k\}$ where $1 \le j \le k \le n-1$. Besides, each item $i$ is associated with several attributes $\mathcal{A}_i = \{a_1, \dots, a_m\}$. For example, a product on Amazon is associated with its title, category, etc. The set of all attributes is denoted the attribute set $\mathcal{A}$, and the number of attributes is denoted as $|\mathcal{A}|$.

Based on the above notations, we now define the task of sequential recommendation. Formally, given the historical behaviors of a user $\{i_1, \dots, i_{n-1}\}$ and the attributes $\mathcal{A}_i$ of each item $i$, the task of sequential recommendation is to predict the next item that the user is likely to interact with at the $n$-th step.

\pageenlarge{2}

\subsection{Last Item Search (LIS) baseline}
\label{sec:lis-baseline}
\begin{figure}
    \centering
    \includegraphics[width=\linewidth]{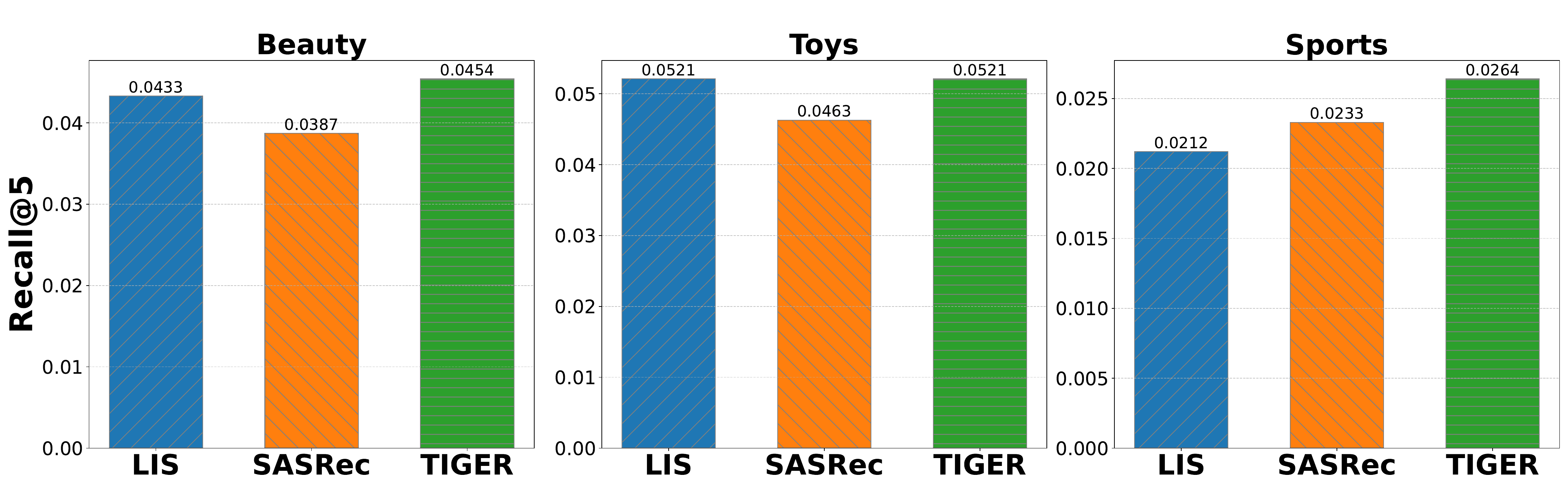}
    \caption{Last-item text-based search (LIS) is a simple, training-free baseline that performs competitively on Recall@5 with modern sequential models like SASRec~\cite{sasrec} and TIGER~\cite{TIGER}. Corresponding tables in Appendix~\ref{app:LIS-Results}}
    \label{fig:LIS-barplot}
\end{figure}
Drawing on the content-based filtering methods discussed in Section \ref{Sec:relatedworks-Content-based}, we began with a preliminary investigation using \textbf{\underline{L}ast \underline{I}tem text \underline{S}earch (LIS)}—a simple, training-free\footnote{We do not consider computing IDF weights in BM25 as training} baseline that retrieves recommendations based on the text content of the last item a user interacted with. Despite its simplicity, this method performs surprisingly well, achieving Recall@5 scores comparable to those of modern sequential recommenders like SASRec and TIGER across the Beauty, Toys, and Sports datasets. Notably, the results shown in Figure~\ref{fig:LIS-barplot} use BM25~\cite{robertson1995okapibm25}, a classic lexical retriever developed over three decades ago.

\subsection{GLoSS pipeline}
\label{sec:Gloss-pipeline}
The GLoSS pipeline is illustrated in Figure~\ref{fig:gloss}. Given a user's item sequence, we:
i) \textit{Serialize} the user's interaction history into natural language text using item metadata (e.g., titles), and \textit{Finetune} a large language model on this data;
ii) \textit{Generate} candidate texts representing the next likely item;
iii) \textit{Retrieve} items from the catalog that closely match these candidate texts.
Here, we present a high-level overview—further implementation details can be found in Section~\ref{sec:implementation-GLoSS}.

\settopmatter{printacmref=false} % Removes citation information below abstract
\renewcommand\footnotetextcopyrightpermission[1]{}
\pagestyle{plain} % removes running headers
\looseness -1  \subsubsection{Serialize and Finetune}
We illustrate our serialization process with a simple example. Suppose a user has interacted with $n = 4$ items. The corresponding training text is constructed using the first $n - 1 = 3$ items and structured as shown below. We then fine-tune a large language model using QLoRA~\cite{qlora} on this serialized training data.

\begin{tcolorbox}[colback=gray!5!white, colframe=gray!75!black, title=Training Text, breakable]
\label{box:training-text}
Below is a customer's purchase history on Amazon, listed in chronological order (earliest to latest).
Each item is represented by the following format: \texttt{Title: <item title>}. Based on this history, predict only one item the customer is most likely to purchase next in the same format.

\textbf{Purchase history:} 

Title: 63cm Long Zipper Beige+pink Wavy Cosplay Hair Wig Rw157  

Title: MapofBeauty Long Wave Curly Hair Wig Full Wig for Women Long (Black)  

\textbf{Next item:}  

Title: 32" 80cm Long Hair Heat Resistant Spiral Curly Cosplay Wig (Red Dark)
\end{tcolorbox}

\begin{tcolorbox}[colback=gray!5!white, colframe=gray!75!black, title=Test Prompt, breakable]
\label{box:test-prompt}
Below is a customer's purchase history on Amazon, listed in chronological order (earliest to latest).
Each item is represented by the following format: \texttt{Title: <item title>}. Based on this history, predict only one item the customer is most likely to purchase next in the same format.

\textbf{Purchase history:}  

Title: 63cm Long Zipper Beige+pink Wavy Cosplay Hair Wig Rw157  

Title: MapofBeauty Long Wave Curly Hair Wig Full Wig for Women Long (Black)  

Title: 32" 80cm Long Hair Heat Resistant Spiral Curly Cosplay Wig (Red Dark)

\textbf{Next item:}
\end{tcolorbox}

\subsubsection{Generate}
To obtain candidate texts for the test target item, we feed the above test prompt, which includes all previous items in the purchase history, into the LLM. The model then continues generation after the ``Next item:\verb|\n|'' token.

\subsubsection{Retrieve}
We construct a dense retrieval index over the item metadata corpus and store it for inference. For each generated text, we perform semantic search by computing the dot product between its embedding and the embeddings of all catalog items, retrieving the closest match.

\pageenlarge{2}
\subsection{Comparison with GPT4Rec and  LlamaRec}
\label{sec:GPT,LlamrecvsGloss}
In this section, we describe the differences between GLoSS and two related methods: GPT4Rec~\cite{GPT4Rec}, which also employs a "generate-then-retrieve" pipeline, and LlamaRec~\cite{Llamarec}, which similarly utilizes LLaMA-series models for recommendation.

First, GPT4Rec~\cite{GPT4Rec} is the most similar approach to ours, and we explicitly acknowledge it as the source of our inspiration. The main differences from GPT4Rec include: (i) the use of semantic search instead of BM25-based lexical matching; (ii) the adoption of the modern LLaMA-3 backbone instead of the older GPT-2; and (iii) the use of QLoRA for parameter-efficient fine-tuning. Collectively, these improvements enable us to achieve higher effectiveness compared to GPT4Rec and reach state-of-the-art results (see Section~\ref{sec:results}).

Second, we are not the first to apply LLaMA to recommendation. Yue et al. proposed LlamaRec~\cite{Llamarec}, which also incorporates LLaMA within the recommendation pipeline. However, its usage differs significantly from ours: LlamaRec employs the LLM for ranking candidates retrieved by a simpler model. In contrast, GLoSS uses LLaMA to retrieve high-quality candidates directly, without additional reranking. Since GLoSS leverages a stronger model at the retrieval stage, it is expected to retrieve a superior set of candidates compared to LlamaRec (i.e., we expect higher recall). However, as we do not perform reranking, ranking-specific metrics such as NDCG may be weaker than those of LlamaRec. The retrieval enhancements of GLoSS and the reranking strengths of LlamaRec are likely complementary. Nevertheless, due to the unavailability of LlamaRec’s trained checkpoints, we leave the investigation of this potential addition for future work.

This summarizes the differences of GLoSS with GPT4Rec and LlamaRec. We now turn to experimental evaluation of GLoSS.

\section{Experiments} \label{sec:experiments}

This section contains experimental evaluation of GLoSS. Section~\ref{ssec:setup}  describes the setup we use for our experiments; Section~\ref{sec:results} contains the results of our experiments.
\pageenlarge{2}

\subsection{Setup}\label{ssec:setup}

\subsubsection{Datasets} 
We perform experiments on three real-world benchmarks from the Amazon Product Reviews dataset \cite{amzn2014}: ``Beauty'' 
(\textbf{Beauty}), ``Toys and Games'' (\textbf{Toys}), ``Sport and Outdoors''  (\textbf{Sports}). These datasets contain user reviews and item metadata collected between May 1996 and July 2014. Details on preprocessing and dataset statistics are provided in Appendix~\ref{app:dataset-stats}.

\subsubsection{Evaluation}
We adopt the widely used leave-last-out split \cite{sasrec,bert4rec,datasplit-strats}, where the last item in the interaction sequence is used for testing and the previous items used for training. 

For evaluation, we utilize two widely adopted top-$k$ metrics: Recall@5 (hit rate) and Normalized Discounted Cumulative Gain at 5 (NDCG@5). Recall@5 assesses whether a relevant item appears within the top-5 predictions, while NDCG@5 also considers the ranked position of the target. To ensure fair and unbiased evaluation, we compute these metrics by scoring all items in the catalog. This avoids the bias of sampled metrics, which can favor popular items~\cite{canamares2020target, rendle-sampled}. While computationally intensive, full-catalog evaluation provides more accurate results, as sampled metrics are inconsistent with exact versions and can misrepresent recommender performance.

\subsubsection{Baselines}
We compare the performance of GLoSS against $10$ existing models, grouped into ID-based and LLM-based. The ID-based are further classified as Classic, Feature-enhanced and Semantic ID-based.
\begin{itemize}
    \item \textbf{Classic ID-based}: SASRec~\cite{sasrec} and BERT4Rec~\cite{bert4rec}.
    \item \textbf{Feature-enhanced ID-based}: FDSA~\cite{Fdsa} and S3Rec~\cite{S3rec}.
    \item \textbf{Semantic ID-based}: TIGER~\cite{TIGER} and ActionPiece~\cite{ActionPiece}.
    \item \textbf{LLM-based}: P5~\cite{P5Rec}, GPT4Rec~\cite{GPT4Rec}, LlamaRec~\cite{Llamarec} \\ and E4SRec~\cite{E4SRecpaper}
\end{itemize}

A detailed description of these models is provided in Appendix~\ref{app:PriorModelsDescription}. Results for the ID-based baselines are taken from publicly available metrics reported in the ActionPiece paper (Appendix G of~\cite{ActionPiece}). For LLM-based recommenders, we report results as published in P5~\cite{P5Rec}, GPT4Rec~\cite{GPT4Rec}, LlamaRec~\cite{Llamarec} and E4SRec~\cite{E4SRecpaper}. Notably, only the Beauty dataset is consistently used across these works, while standard benchmarks like Toys and Sports are not reported for GPT4Rec and LlamaRec. 
We do not reproduce their models on these datasets due to (i) the lack of open-source code models (e.g., GPT4Rec) and (ii) compute constraints\footnote{Our experiments are limited to a single RTX A5000 GPU.}. This aligns with common practice in the LLM-based recommender literature, where both training and inference are more resource-intensive than ID-based methods. For instance, LlamaRec~\cite{Llamarec} reports results only on the Beauty dataset for this reason.

\subsubsection{Implementation Details for GLoSS}
\label{sec:implementation-GLoSS}
\paragraph{Finetuning}  
The LLMs used in our GLoSS pipeline are from the LLaMA-3 model family \cite{llama3}. We use models with increasing parameter sizes: LLaMA 3.2 (1B, 3B) and LLaMA 3.1 (8B). We limit the maximum context length to 1024 tokens, truncating from the left if necessary. The models are finetuned using QLoRA \cite{qlora} via the Unsloth library \cite{unsloth}, with 4-bit quantization and LoRA rank and $\alpha$ set to $16$. This results in 11M, 24M, and 42M trainable adapter parameters for the 1B, 3B, and 8B models, respectively, enabling training on a single RTX A5000 GPU with 24GB memory.

\pageenlarge{2}
We train for $10$ epochs. Our batch size and gradient accumulation are $4$ each, yielding an effective batch size of $16$. Optimization is performed using AdamW with a learning rate of $0.0001$, a linear schedule with $300$ warm-up steps, and a weight decay of $0.01$. Early stopping is based on validation loss, with a patience of $3$ epochs.
\paragraph{Generation}
For text generation, we use beam search decoding with 5 beams, producing 5 candidate texts per user. The \texttt{max\_new\_tokens} i.e., the tokens generated after ``Next item:'') is set to $50$.
\paragraph{Retrieval}  
For retrieval, we use the \texttt{e5-small-v2} encoder\footnote{\url{https://huggingface.co/intfloat/e5-small-v2}} to index the item catalog~\cite{wang2024textembsE5}. This model has 33M parameters and produces embeddings of dimension $d = 384$. The dense index is constructed using the \texttt{retriv} package\footnote{\url{https://github.com/AmenRa/retriv}}.

We denote each GLoSS variant with the suffix \textbf{GLoSS-1B}, \textbf{GLoSS-3B}, or \textbf{GLoSS-8B}, indicating the LLaMA-3 model size used for query generation. Unless otherwise specified, experiments in Sections~\ref{sec:Overall-compare} and~\ref{Sec:RQ-User-segments} use \texttt{e5-small-v2} as the retriever. In Section~\ref{sec:RQ-DenseRetSize}, we compare against a larger dense retriever (\texttt{e5-base-v2}) and a sparse BM25 retriever to assess the importance of dense retrieval and whether encoder size impacts performance. The code, datasets and checkpoints are open-sourced at \url{https://github.com/krishnacharya/GLoSS}.

\subsection{Results}\label{sec:results}
In this section, we evaluate the performance of our GLoSS pipeline on the three Amazon datasets—Beauty, Toys, and Sports. Our evaluation is guided by the following key research questions, which aim to measure how GLoSS compares to existing sequential recommenders (both ID-based and LLM-based), and to examine the impact of dense retrieval, and user interaction length:
\vspace{1\baselineskip}

\noindent\textbf{\small RQ1} Does GLoSS outperform existing sequential recommenders, including both ID-based and LLM-based baselines?\\
\textbf{\small RQ2} What is the effect of using dense retrieval, and does query encoder size matter?\\
\textbf{\small RQ3} How does GLoSS perform across users with varying interaction lengths, and can it effectively handle cold-start users?\\
We now answer these research questions in turn.

\subsubsection{RQ1: Does GLoSS outperform existing baselines?} 
\label{sec:Overall-compare}

We train three GLoSS variants—GLoSS-1B, GLoSS-3B, and GLoSS-8B, corresponding to different sizes of the LLaMA-3 backbone. In Table~\ref{tab:ID-basedvsGLoSS}, we compare these against six ID-based baselines: SASRec, BERT4Rec, FDSA, S3Rec, TIGER, and ActionPiece. GLoSS-8B consistently outperforms all ID-based methods, achieving state-of-the-art performance. The improvements in Recall@5 are substantial, with gains of 33.27\%, 52.78\%, and 15.19\% on the Beauty, Toys, and Sports datasets respectively. Similarly, we observe NDCG@5 improvements of 30\%, 42.59\%, and 16.1\%. We also find that performance improves with larger LLaMA model sizes, which is expected given the enhanced generation quality of larger language models.

\looseness -1 In Table~\ref{tab:LLMbasevsGLoSS}, we compare GLoSS against four LLM-based recommenders: P5, GPT4Rec, LlamaRec, and E4SRec. GLoSS achieves higher Recall@5 across all datasets—showing gains of 4.29\%, 22.84\%, and 29.54\% on Beauty, Toys, and Sports, respectively. While GLoSS is competitive in NDCG@5, it does not surpass LlamaRec, which benefits from an additional reranking stage. As discussed in Section~\ref{sec:GPT,LlamrecvsGloss}, LlamaRec retrieves candidates using a simpler model and applies an LLM for reranking, optimizing ranking metrics. In contrast, GLoSS uses LLaMA directly for retrieval, yielding high-quality candidates and strong recall. The two approaches are likely complementary, but we leave exploration of a combined system to future work due to the unavailability of LlamaRec’s checkpoints.

\looseness -1 We also note that P5’s reported results on the Toys dataset are likely optimistic; a recent reproduction study~\cite{P5Repro} reports significantly lower scores. However, to ensure a conservative and robust comparison, we retain the numbers from the original P5 paper~\cite{P5Rec}.

These results support that GLoSS achieves state-of-the-art performance among both ID-based and LLM-based recommenders, demonstrating the efficacy of integrating dense retrieval with strong LLM generation backbones.

\begin{table*}[!h]
\centering
\caption{Performance of GLoSS compared to ID-based sequential recommenders. The best overall metric is shown in \textbf{bold}, and the second best is \underline{underlined}. The $\dagger$ symbol indicates the best-performing baseline (excluding our models). GLoSS-8B achieves the best performance across all datasets, with percentage improvements over $\dagger$ reported in brackets. \label{tab:ID-basedvsGLoSS}}
\begin{tabular}{l|cc|cc|cc}
\hline
\multirow{2}{*}{Model} & \multicolumn{2}{c|}{Beauty} & \multicolumn{2}{c|}{Toys} & \multicolumn{2}{c}{Sports} \\
\cline{2-7}
 & Recall@5 & NDCG@5 & Recall@5 & NDCG@5 & Recall@5 & NDCG@5 \\
SASRec & 0.0387 & 0.0249 & 0.0463 & 0.0306 & 0.0233 & 0.0154 \\
BERT4Rec & 0.0203 & 0.0124 & 0.0116 & 0.0071 & 0.0115 & 0.0075 \\
FDSA & 0.0267 & 0.0163 & 0.0228 & 0.0140 & 0.0182 & 0.0122 \\
S3Rec & 0.0387 & 0.0244 & 0.0443 & 0.0294 & 0.0251 & 0.0161 \\
TIGER & 0.0454 & 0.0321 & 0.0521$\dagger$ & 0.0371$\dagger$ & 0.0264 & 0.0181 \\
ActionPiece & 0.0511$\dagger$ & 0.0340$\dagger$ & N/A & N/A & \underline{0.0316}$\dagger$ & \underline{0.0205}$\dagger$ \\
\hline
GLoSS-1B  & 0.0456 & 0.0297 & 0.0677 & 0.0441 & 0.0226 & 0.0145 \\
GLoSS-3B & \underline{0.0653} & \underline{0.0423} & \underline{0.0728} & \underline{0.0478} & 0.0294 & 0.0188 \\
GLoSS-8B & \textbf{0.0681} (+33.27\%) & \textbf{0.0442} (+30.00\%) & \textbf{0.0796} (+52.78\%) & \textbf{0.0529} (+42.59\%) & \textbf{0.0364} (+15.19\%) & \textbf{0.0238} (+16.10\%) \\
\hline
\end{tabular}
\end{table*}

\begin{table*}[!h]
\centering
\caption{Performance of GLoSS compared to LLM based recommenders. The best metric overall is in \textbf{bold} and second best is \underline{underlined}. The $\dagger$ symbol indicates the best-performing baseline (excluding our models). GLoSS-8B achieves the highest Recall@5  across all datasets.\label{tab:LLMbasevsGLoSS}}
\begin{tabular}{l|cc|cc|cc}
\hline
\multirow{2}{*}{Model} & \multicolumn{2}{c|}{Beauty} & \multicolumn{2}{c|}{Toys} & \multicolumn{2}{c}{Sports} \\
\cline{2-7}
 & Recall@5 & NDCG@5 & Recall@5 & NDCG@5 & Recall@5 & NDCG@5 \\
\hline
P5 & 0.0503 & 0.0370 & 0.0648$\dagger$ & \textbf{0.0567}$\dagger$ & 0.0272 & 0.0169\\
GPT4Rec & 0.0653$\dagger$ & N/A & N/A & N/A  & N/A  & N/A  \\
LlamaRec & 0.0648 & \textbf{0.0450}$\dagger$ & N/A & N/A & N/A & N/A \\
E4SRec & 0.0525 & 0.0360 & 0.0566 & 0.0405 & 0.0281$\dagger$ & \underline{0.0196}$\dagger$ \\
\hline
GLoSS-1B & 0.0456 & 0.0297 & 0.0677 & 0.0441 & 0.0226 & 0.0145 \\
GLoSS-3B & \underline{0.0653} & 0.0423 & \underline{0.0728} & 0.0478 & \underline{0.0294} & 0.0188 \\
GLoSS-8B & \textbf{0.0681} (+4.29\%) & \underline{0.0442} (-1.77\%) & \textbf{0.0796} (+22.84\%) & \underline{0.0529} (-6.70\%) & \textbf{0.0364} (+29.54\%) & \textbf{0.0238} (+21.43\%) \\
\hline
\end{tabular}
\end{table*}

\subsubsection{RQ2 How important is dense retrieval and does encoder size matter?} \label{sec:RQ-DenseRetSize}
We investigate the impact of dense retrieval by directly comparing it to traditional lexical retrieval using BM25. Specifically, we evaluate two dense retrievers of increasing model capacity: \texttt{e5-small-v2} (33M parameters, embedding dimension $d=384$) and \texttt{e5-base-v2} (109M parameters, $d=768$). As shown in Table~\ref{tab:bm25-vs-denseretrievers}, dense retrieval consistently outperforms BM25, with substantial gains in NDCG@5, reaching double digits. Improvements in Recall@5, while generally smaller in magnitude, still reflect the value of semantic representations in identifying relevant items beyond lexical matches. Interestingly, a larger encoder does not always yield better performance; for example, on the Beauty dataset, \texttt{e5-small-v2} outperforms \texttt{e5-base-v2} when used with the 1B and 8B GLoSS variants.

\begin{table*}[!h]
\centering
\caption{Comparison of different retrievers on generated texts. Dense retrieval outperforms BM25, showing substantial gains in NDCG@5 and moderate improvements in Recall@5. Brackets indicate percentage improvements over BM25.}
\begin{tabular}{l|l|cc|cc|cc}
\hline
\multicolumn{2}{c|}{\textbf{Model Configuration}} & \multicolumn{2}{c|}{\textbf{Beauty}} & \multicolumn{2}{c|}{\textbf{Toys}} & \multicolumn{2}{c}{\textbf{Sports}} \\
\cline{3-8}
\multicolumn{2}{c|}{} & \textbf{Recall@5} & \textbf{NDCG@5} & \textbf{Recall@5} & \textbf{NDCG@5} & \textbf{Recall@5} & \textbf{NDCG@5} \\
\hline
\textbf{1B} & \textbf{BM25} & 0.0453 & 0.0269 & 0.0666 & 0.0401 & 0.0221 & 0.0133 \\
            & \textbf{e5-small-v2} & 0.0456 (+0.66\%) & 0.0297 (+10.41\%) & 0.0677 (+1.65\%) & 0.0441 (+9.98\%) & 0.0226 (+2.26\%) & 0.0145 (+9.02\%) \\
            & \textbf{e5-base-v2} & 0.0458 (+1.10\%) & 0.0295 (+9.67\%) & 0.0675 (+1.35\%) & 0.0443 (+10.47\%) & 0.0228 (+3.17\%) & 0.0139 (+4.51\%) \\
\hline
\textbf{3B} & \textbf{BM25} & 0.0652 & 0.0400 & 0.0720 & 0.0436 & 0.0288 & 0.0178 \\
            & \textbf{e5-small-v2} & 0.0653 (+0.15\%) & 0.0423 (+5.75\%) & 0.0728 (+1.11\%) & 0.0478 (+9.63\%) & 0.0294 (+2.08\%) & 0.0188 (+5.62\%) \\
            & \textbf{e5-base-v2} & 0.0652 (+0.00\%) & 0.0425 (+6.25\%) & 0.0728 (+1.11\%) & 0.0474 (+8.72\%) & 0.0295 (+2.43\%) & 0.0182 (+2.25\%) \\
\hline
\textbf{8B} & \textbf{BM25} & 0.0681 & 0.0423 & 0.0784 & 0.0472 & 0.0359 & 0.0218 \\
            & \textbf{e5-small-v2} & 0.0681 (+0.00\%) & 0.0442 (+4.49\%) & 0.0796 (+1.53\%) & 0.0529 (+12.08\%) & 0.0364 (+1.39\%) & 0.0238 (+9.17\%) \\
            & \textbf{e5-base-v2} & 0.0683 (+0.29\%) & 0.0435 (+2.84\%) & 0.0790 (+0.77\%) & 0.0526 (+11.44\%) & 0.0365 (+1.67\%) & 0.0228 (+4.59\%) \\
\hline
\end{tabular}
\label{tab:bm25-vs-denseretrievers}
\end{table*}

\subsubsection{RQ3: How does GLoSS perform across users with varying interaction histories, and can it effectively handle cold-start users?}
\label{Sec:RQ-User-segments}

We categorize users into three groups based on the length of their interaction sequences. Given a user history $U_i = \{i_1, i_2, ..., i_n\}$, $|U_i| = n$, users are classified as:

\begin{itemize}
    \item \textbf{Short-sequence (cold-start) users:} $n \le l$
    \item \textbf{Medium-sequence (regular) users:} $l < n \le h$
    \item \textbf{Long-sequence (power) users:} $n > h$
\end{itemize}

The choices for the $l, h$ thresholds and the corresponding group sizes are available in Appendix~\ref{app:userseg-results}. This segmentation allows us to evaluate how model performance varies with the size of user history. Having a model that recommends well with limited interactions is critical for new customer retention \cite{wang2021colduserseq,cold-both-ui}, and we note that similar user segment-wise evaluation has been proposed in prior work~\cite{DRO-TTGoogle, minimaxseqrec}

In the interest of space, Tables~\ref{tab:recall5_user_segment_metrics} and~\ref{tab:ndcg5_user_segment_metrics} reporting user segment performance are presented in Appendix~\ref{app:userseg-seqthresholds}. These show that GLoSS achieves high Recall@5 and NDCG@5 for cold-start users. In fact, in the Toys and Sports datasets, users with short interaction histories achieve the highest performance. In contrast, the Beauty dataset shows the opposite trend: performance improves with longer histories. This suggests that in Toys and Sports, recent interactions may provide more distinctive signals, whereas in Beauty, longer histories offer richer context. Similar trends are observed with the Last-item text search baseline in Appendix~\ref{app:LIS-Results}, indicating that this is a dataset-specific rather than model-specific effect.

In any case, GLoSS shows robust performance across datasets, remaining effective in both sparse and interaction-rich scenarios.

\section{Conclusion} \label{sec:concl}
In this paper, we introduced GLoSS (Generative Low-rank language model with Semantic Search), a novel generative recommendation framework. GLoSS distinguishes itself from prior approaches like GPT4Rec by replacing lexical BM25 retrieval with semantic search, leading to more accurate and context-aware item retrieval. For efficient query generation, GLoSS leverages 4-bit quantized LLaMA-3 models fine-tuned via LoRA, enabling efficient training and inference without sacrificing quality.

Our comprehensive experiments on three real-world Amazon review datasets—Beauty, Toys, and Sports—demonstrate that GLoSS consistently achieves state-of-the-art performance. It significantly outperforms ID-based baselines in both Recall@5 and NDCG@5, and surpasses existing LLM-based recommenders on Recall@5. While GLoSS shows competitive but sometimes slightly lower NDCG scores compared to LLamaRec on Beauty and P5Rec on Toys, a promising future direction involves incorporating LLMs as judges for reranking; this could further improve NDCG and overall recommendation quality. Beyond overall performance, GLoSS demonstrates high performance robustness across user segments, with notably high performance for cold-start users on the Sports and Toys datasets.

\begin{acks}
The authors gratefully acknowledge the compute resources used in these experiments, provided by Simian—a GPU workstation set up by Prof. Jacob Abernethy and maintained by PhD student Tyler LaBonte. Prof. Ziani's research was supported by NSF CAREER Award IIS-2336236. 
\end{acks}

\newpage

\bibliographystyle{ACM-Reference-Format}
\bibliography{refs}

\appendix
\section{Discussion and Limitations}
Here, we discuss some limitations of LLM-based recommenders. Compared to ID-based models, larger LLMs such as the 8B LLaMA model used in our experiments take longer to train and require more VRAM during both fine-tuning and generation. We hope that our open-source codebase, which uses quantized LoRA fine-tuning (via \texttt{Unsloth}) and integrates paged attention (with \texttt{vLLM}), can help the community address these common challenges.

Additionally, large language models are typically trained on vast internet-scale corpora that may include portions of open datasets used for benchmarking. Recent work studying memorization on Movielens-1M \cite{llm-Movielensmemorize} shows that LLaMA-3 models can memorize datasets like Movielens, with larger models exhibiting higher memorization rates. For example, the memorization rate reported for LLaMA-3.1 405B is 12.9\%, while for LLaMA-3.1 8B it is lower at 5.82\% percent.
Since we evaluate GLoSS on Amazon datasets and use LLaMA-3 models (1B, 3B, and 8B) that exhibit relatively low memorization rates—at least in the case of Movielens—we consider the memorization risk in our setup to be low.

\section{Dataset statistics}
\label{app:dataset-stats}
We conduct experiments on three real-world datasets from the Amazon Product Reviews corpus~\cite{amzn2014}, specifically the ``Beauty'' (\textbf{Beauty}), ``Toys and Games'' (\textbf{Toys}), and ``Sports and Outdoors'' (\textbf{Sports}) categories. These datasets contain user reviews and item metadata collected between May 1996 and July 2014. We use the official 5-core subset~\cite{mcauley2014v1, mcauley2014v2}, which retains users with at least five reviews.\footnote{Available at \url{https://jmcauley.ucsd.edu/data/amazon/index_2014.html}. All compared methods also adopt 5-core filtering. For models applying their own filtering (e.g., TIGER~\cite{TIGER}), the number of interactions typically differs by few 100s from Table~\ref{tab:dataset_summary}, which we believe does not affect metrics.} User--item interactions are grouped by user and sorted chronologically based on interaction timestamps. For evaluation, we adopt the standard leave-last-item-out split~\cite{sasrec,bert4rec,datasplit-strats}. A summary of dataset statistics is provided in Table~\ref{tab:dataset_summary}.

\begin{table}[h!]
    \centering
    \caption{Statistics for the three Amazon datasets}
    \label{tab:dataset_summary}
    \begin{tabular}{|l|c|c|c|c|}
        \hline
        \textbf{Dataset} & \makecell{\textbf{\#Users}} & \makecell{\textbf{\#Items}} & \makecell{\textbf{\#Interactions}} & \makecell{\textbf{Mean Seq} \\ \textbf{Length}} \\
        \hline
        Beauty & 22,363 & 12,094 & 198,371 & 8.87 \\
        Toys & 19,406 & 11,865 & 166,757 & 8.59 \\
        Sports & 35,597 & 18,267 & 295,091 & 8.29 \\
        \hline
    \end{tabular}
\end{table}
\vspace{-2\baselineskip}

\section{Evaluation Across User Segments}
\subsection{Thresholding by Sequence Length}
\label{app:userseg-seqthresholds}
The threshold choices for segmenting users into cold-start, regular, and power user groups are summarized in Table~\ref{tab:user_segments_only}. We set the lower threshold $l=5$ across all three datasets, while the upper threshold $h$ is set to 14, 13, and 13 for the Beauty, Toys, and Sports datasets, respectively. These thresholds result in approximately 10\% power users, 60\% regular users, and 30\% cold-start users.

\begin{table}[h!]
\centering
\caption{User Distribution Across Segments Based on Sequence Length Thresholds}
\label{tab:user_segments_only}
\begin{tabular}{l|l|c}
\hline
\textbf{Dataset} & \textbf{Sequence Length} & \textbf{User Split} \\
 & \textbf{Segment} & \textbf{Percentage (\%)} \\
\hline
\multirow{3}{*}{\textbf{Beauty}}
 & $\le 5$ (Cold-Start) & 32.11 \\
 & (5, 14] (Regular) & 58.11 \\
 & $> 14$ (Power) & 9.78 \\
\hline
\multirow{3}{*}{\textbf{Toys}}
 & $\le 5$ (Cold-Start) & 34.52 \\
 & (5, 13] (Regular) & 56.34 \\
 & $> 13$ (Power) & 9.14 \\
\hline
\multirow{3}{*}{\textbf{Sports}}
 & $\le 5$ (Cold-Start) & 32.60 \\
 & (5, 13] (Regular) & 59.00 \\
 & $> 13$ (Power) & 8.40 \\
\hline
\end{tabular}
\end{table}

\subsection{Performance Across Segments}

\label{app:userseg-results}
Tables~\ref{tab:recall5_user_segment_metrics} and~\ref{tab:ndcg5_user_segment_metrics} report performance across user segments for Recall@5 and NDCG@5, respectively. GLoSS achieves consistently high values for both metrics across all 3 segments.

\begin{table}[H]
\centering
\caption{Recall@5 for the three user segments. The \textbf{*} symbol indicates the user segement with the highest performance for a given dataset and model.}
\resizebox{\linewidth}{!}{
\label{tab:recall5_user_segment_metrics}
\begin{tabular}{ll|c|ccc} % Dataset | Model | Overall | Cold-Start | Regular | Power Users
\toprule
\textbf{Dataset} & \textbf{Model} & \textbf{Overall} & \textbf{Short} & \textbf{Med} & \textbf{Long} \\
\midrule
\multirow{3}{*}{\textbf{Beauty}} & GLoSS-1B & 0.0456 & 0.0439 & 0.0449 & 0.0549* \\
& GLoSS-3B & 0.0653 & 0.0575 & 0.0626 & 0.1065* \\
& GLoSS-8B & 0.0681 & 0.0618 & 0.0640 & 0.1129* \\
\midrule
\multirow{3}{*}{\textbf{Toys}} & GLoSS-1B & 0.0677 & 0.0755* & 0.0640 & 0.0617 \\
& GLoSS-3B & 0.0728 & 0.0809* & 0.0689 & 0.0660 \\
& GLoSS-8B & 0.0796 & 0.0861* & 0.0774 & 0.0694 \\
\midrule
\multirow{3}{*}{\textbf{Sports}} & GLoSS-1B & 0.0226 & 0.0275* & 0.0211 & 0.0156 \\
& GLoSS-3B & 0.0294 & 0.0333* & 0.0280 & 0.0250 \\
& GLoSS-8B & 0.0364 & 0.0401* & 0.0349 & 0.0327 \\
\bottomrule
\end{tabular}
}
\end{table}

\begin{table}[!h]
\centering
\caption{NDCG@5 for the three user segments. The \textbf{*} symbol indicates the user segment with the highest performance for a given dataset and model.}
\label{tab:ndcg5_user_segment_metrics}
\resizebox{\linewidth}{!}{
\begin{tabular}{ll|c|ccc} % Dataset | Model | Overall | Cold-Start | Regular | Power Users
\toprule
\textbf{Dataset} & \textbf{Model} & \textbf{Overall} & \textbf{Short} & \textbf{Med} & \textbf{Long} \\
\midrule
\multirow{3}{*}{\textbf{Beauty}} & GLoSS-1B & 0.0297 & 0.0272 & 0.0298 & 0.0370* \\
& GLoSS-3B & 0.0423 & 0.0362 & 0.0406 & 0.0724* \\
& GLoSS-8B & 0.0442 & 0.0398 & 0.0414 & 0.0750* \\
\midrule
\multirow{3}{*}{\textbf{Toys}} & GLoSS-1B & 0.0441 & 0.0515* & 0.0407 & 0.0375 \\
& GLoSS-3B & 0.0478 & 0.0534* & 0.0451 & 0.0430 \\
& GLoSS-8B & 0.0529 & 0.0583* & 0.0510 & 0.0455 \\
\midrule
\multirow{3}{*}{\textbf{Sports}} & GLoSS-1B & 0.0145 & 0.0176* & 0.0135 & 0.0102 \\
& GLoSS-3B & 0.0188 & 0.0213* & 0.0179 & 0.0163 \\
& GLoSS-8B & 0.0238 & 0.0262* & 0.0226 & 0.0226 \\
\bottomrule
\end{tabular}}
\end{table}

\section{Detailed description of prior ID-based and LLM based recommenders}
\label{app:PriorModelsDescription}
We compare GLoSS with the following ID-based and LLM based models 
\subsection{Classic ID-based}
\begin{itemize}
    \item SASRec \cite{sasrec} Represent each item by a unique item ID, it encodes a user's item ID sequence with a self-attentive Transformer decoder, the model is trained using BCE loss.
    \item BERT4Rec \cite{bert4rec} also represents each item by unique item ID, but it encodes user sequences with a bidirectional transform encode (like BERT \cite{bert4rec}). The model is trained using a masked prediction objective and the cross entropy loss 
\end{itemize}

\subsection{Feature+ID-based}
\begin{itemize}
    \item FDSA \cite{Fdsa} Integrates item feature embeddings with vanilla attention layers to obtain feature representations. It then processes item ID and feature sequences seperately through self-attention blocks
    \item S3Rec \cite{S3rec} uses self supervised pre-training to capture correlation between item feature and item ID, these checkpoints are then loaded and finetuned for next-item prediction using only item IDs.
\end{itemize}

\subsection{Semantic ID-based}
\begin{itemize}
    \item TIGER \cite{TIGER} encodes text features and quantizes them into semantic IDs using RQ-VAE. The model is then trained to predict the next semantic ID and uses beam search during inference.
    \item ActionPiece~\cite{ActionPiece} incorporates context by building a tokenizer that maps sets of item features to actions, generating tokens based on the co-occurrence patterns of these features.
\end{itemize}

\subsection{LLM based}
\begin{itemize}
    \item P5Rec \cite{P5Rec} T5 \cite{T5paper} as a unified backbone to support multiple tasks, including sequential recommendation, rating prediction, and review summarization.
    \item GPT4Rec \cite{GPT4Rec} leverages GPT-2 to generate multiple queries per user, which are then matched to items using a BM25 lexical retriever
    \item LlamaRec~\cite{Llamarec} employs Llama-2 7B as a reranker over candidates retrieved via an LRU-based recommender~\cite{LRURec}
    \item E4SRec~\cite{E4SRecpaper} operates in two phases. First, it trains a SASRec using only item ID information. Then, it instruction-tunes a LLaMA-2 13B model, where the serialized user history is used as input, and at each position i, the input token consists of a concatenation of the item's text embedding and its corresponding SASRec-derived item embedding.

\end{itemize}

\section{Additional Results}
\subsection{Results with Last item text search (LIS)}
\label{app:LIS-Results}

\looseness -1 Building on the content-based filtering methods discussed in Section~\ref{Sec:relatedworks-Content-based}, we conducted a preliminary investigation using \textbf{\underline{L}ast \underline{I}tem text \underline{S}earch (LIS)}, a training-free baseline that retrieves recommendations based on the text of the last item a user interacted with.

Figure~\ref{fig:LIS-barplot} in Section~\ref{sec:lis-baseline} showed LIS with the BM25~\cite{robertson1995okapibm25} retriever; we also evaluated dense encoders (e5-small-v2 and e5-base-v2) to enable semantic similarity-based last item search.

Despite its simplicity, LIS with all three retrievers performs competitively, achieving Recall@5 and NDCG@5 scores comparable to modern sequential recommenders like SASRec and TIGER across the Beauty, Toys, and Sports datasets. Full results are provided in Table~\ref{tab:LIS-comparison}. 

\begin{table}[h]
\centering
\caption{Metrics for Last Item Search (LIS), SASRec, and TIGER. Best in bold, second best underlined}
\vspace{-1\baselineskip}
\label{tab:LIS-comparison}
\resizebox{\linewidth}{!}{
\begin{tabular}{llcc}
\toprule
\textbf{Dataset} & \textbf{Model} & \textbf{Recall@5} & \textbf{NDCG@5} \\
\midrule
\multirow{5}{*}{\textbf{Beauty}} 
& LIS-BM25 & 0.0433 & 0.0232 \\
& LIS-e5-small-v2 & 0.0418 & 0.0224 \\
& LIS-e5-base-v2 & \underline{0.0421} & 0.0226 \\
& SASRec & 0.0387 & \underline{0.0249} \\
& TIGER & \textbf{0.0454} & \textbf{0.0321} \\
\midrule
\multirow{5}{*}{\textbf{Toys}} 
& LIS-BM25 & 0.0521 & 0.0267 \\
& LIS-e5-small-v2 & \underline{0.0553} & 0.0288 \\
& LIS-e5-base-v2 & \textbf{0.0572} & 0.0295 \\
& SASRec & 0.0463 & \underline{0.0306} \\
& TIGER & 0.0521 & \textbf{0.0371} \\
\midrule
\multirow{5}{*}{\textbf{Sports}} 
& LIS-BM25 & 0.0212 & 0.0110 \\
& LIS-e5-small-v2 & 0.0212 & 0.0112 \\
& LIS-e5-base-v2 & 0.0216 & 0.0113 \\
& SASRec & \underline{0.0233} & \underline{0.0154} \\
& TIGER & \textbf{0.0264} & \textbf{0.0181} \\
\bottomrule
\end{tabular}
}
\end{table}

\section{Generated text samples}
Here, we present an example of the $5$ generated texts from the finetuned models. The outputs demonstrate that the LLMs produce well-structured candidate texts, effectively capturing the context of the seen items' texts. The target item's text is highlighted in blue.
\begin{tcolorbox}[colback=gray!5!white, colframe=gray!75!black, title=Sample generated texts with the finetuned Llama-3 models]

\textbf{Target ASIN Text:}

\textcolor{blue}{Title: Sigma F80 - Flat Kabuki TM}

\textbf{Seen ASINs Texts:}

Title: SKINFOOD Salmon Darkcircle Concealer Cream \#1 Blooming Light Beige (Whitening Care) 10g

Title: SKIN79 Hot Pink Super Plus Beblesh Balm 40g

Title: Laneige Water Sleeping Pack EX 80ml

Title: Sigma E65 - Small Angle

\hrule % Horizontal line for separation
\vspace{1mm}
\textbf{Generated Texts with Finetuned Llama1b:}
\vspace{1mm}

Title: SKINFOOD Salmon Darkcircle Concealer Cream \#1 Blooming Light Beige (Whitening Care) 10g

Title: SKIN79 Super Plus Beblesh Balm Triple Functions - VIP Gold 40g

Title: Sigma F80 - Flat Kabuki TM

Title: SKINFOOD Salmon Darkcircle Concealer Cream \#1 Blooming Light-beige (Whitening Care) 10g

Title: SKINFOOD Salmon Darkcircle Concealer Cream \#1 blooming light beige (Whitening Care) 10g

\hrule % Horizontal line for separation
\vspace{1mm}
\textbf{Generated Texts with Finetuned Llama3b:}
\vspace{1mm}

Title: Sigma E75 - Angled Brow

Title: Etude House Precious Mineral BB Cream Bright Fit SPF30/PA++ \#N02 Light Beige

Title: Etude House Precious Mineral BB Cream Bright Fit SPF30/PA++ \#W13 Natural Beige

Title: SKIN79 BB Cleanser with Skin Detox Effect

Title: Etude House Precious Mineral BB Cream Bright Fit SPF30/PA++ \#W24 Honey Beige

\hrule % Horizontal line for separation

\vspace{1mm}
\textbf{Generated Texts with Finetuned Llama-8b:}
\vspace{1mm}

Title: SKINFOOD Peach Sake Pore BB Cream \#2(SPF20/PA+) 30ml

Title: SKIN79 Super Plus Triple Functions BB Vital Cream 40g

Title: Etude House Wonder Pore Freshner 500ml

Title: Sigma E75 - Angled Brow

Title: Sigma F80 - Flat Kabuki TM

\normalsize % Revert to normal font size after the box
\end{tcolorbox}

\end{document}